\documentclass[fleqn,10pt]{wlscirep}

\usepackage{orcidlink}

\usepackage[utf8]{inputenc}
\usepackage[T1]{fontenc}
\usepackage{physics}
\usepackage{tikz, tcolorbox}
\usepackage{xcolor}
\usepackage[symbol]{footmisc}
\usepackage{hyperref}

\usepackage{blochsphere}
\usepackage{subcaption}

\title{Intro2QC: An Approachable Introduction to Quantum Computing for STEM Education in High Schools}
\author[1]{Jaimie A. Greasley\orcidlink{0000-0001-7821-6053}}
\author[2,1]{Thomas E. Baker\orcidlink{0000-0002-3142-0767}}
\affil[1]{Department of Chemistry, University of Victoria, Victoria, British Columbia V8P 5C2, Canada}
\affil[2]{Department of Physics \& Astronomy, University of Victoria, Victoria, British Columbia V8P 5C2, Canada}


\begin{abstract} % 279 words
In recent years, quantum computing has gained strong traction on the global stage. Pioneering developments in quantum engineering, algorithms and software reinforce the view that quantum computation will eventually catalyze transformative technological advancements across diverse sectors. In anticipation of the future impact of quantum computing, many governments, universities and tech organizations have invested heavily into advancing quantum science, boosting quantum readiness, and building a diverse, talented workforce through quantum education initiatives. Specialized university programs, public awareness campaigns and STEM outreach activities all serve a critical function in fostering a deeper understanding of quantum physics and quantum technologies among non-experts. At the level of high school, a major vision is for these initiatives to reshape the teaching of 20th century physics in STEM classrooms, and inspire the future generation of quantum scientists and professionals. In the current paper, we are presenting the materials we have created for an outreach workshop introducing quantum computing to high school students in an approachable and engaging way. Intro2QC is an exploration of quantum computing's relevance to solving real-world challenges by relating how the unintuitive behaviours of quantum systems can be useful in providing computational advantage. The workshop materials include an online interactive quantum programming tutorial hosted on the Intro2QC website, along with the full lesson plan for STEM instructors and the presentation slides available on the Github repository. This single-session 90-minute workshop has been delivered to approximately 160 students in multiple school groups from Grades 9-12 in British Columbia Canada in the highlight of the International Year of Quantum Science and Technology. 

\end{abstract}

\begin{document}
\flushbottom
\maketitle

\thispagestyle{empty}

\section{Introduction} %862 words right now
%P1 - 179 words 
Over the past century, technologies based on quantum physical phenomena have profoundly shaped the modern world. From magnetic resonance imaging (MRI), to lasers, GPS navigation with atomic clocks, and semiconductor technologies like transistors and LEDs; the role of quantum physics in the operation of today's electronics is far more extensive than commonly perceived. Still, there has been heightened interest surrounding quantum technologies in the last 2-3 decades, as scientists seek to harness the full power of quantum properties through finer control and direct manipulation of quantum systems. Scientific and engineering advancements have revealed a remarkable potential for quantum-based communication networks, sensing and metrology, simulation and computation, thus marking the onset of a "second quantum revolution" \cite{dowling2003quantum, acin2018quantum}. Amid these new developments, and in celebration of 100 years since the formulation of quantum mechanics, the United Nations announced 2025 as the International Year of Quantum Science and Technology (IYQ) \cite{IYQ2025}. Organized IYQ events have been hosted by numerous institutes and members of the quantum community to raise public awareness of quantum science worldwide, and provide a platform for discussions and knowledge exchange between academia, industry and governing bodies\cite{bongs2025celebrating}.

%P2 - Quantum algorithms - 271 words
Quantum computing is an exciting emergent technology that promises to transform problem-solving in domains that are intractable for classical computing. The first notions of a quantum mechanical computer began to take form in the 1980s owing to some discussions amongst physicists on the intimate relationship between physics and computation \cite{NatRevPhys2022}. These ideas were first documented by Yuri Manin and Paul Benioff around 1980 \cite{manin2007mathematics, benioff1980computer}, but further developed and popularized through a keynote speech by physicist Richard Feynman in 1981  \cite{feynman1982simulating}. Feynman's motivation was rooted in the inefficiencies of classical computing for simulating quantum physics. He is cited as giving this remark at the end of his lecture: "... because nature isn't
classical ... and if you want to make a simulation of nature, you'd better make it quantum mechanical." 

Since that time, the perception of quantum computation as a tool useful only for quantum mechanics has been completely dispelled. David Deutsch put forth the concept of a \textit{universal quantum computer} in 1985, and subsequently developed the quantum circuit model which established a framework for quantum computation on a broader range of computational tasks \cite{deutsch1989quantum, nielsen2010quantum}. Deutsch's algorithm, claimed as the world's first quantum algorithm, showed an exponential speedup over classical computing for a decision problem\cite{deutsch1985quantum, deutsch1992rapid}\footnote[1]{In computer science, a decision problem is one that can be constructed as a "yes/no" or "true/false" outcome for the set of all possible inputs. To determine the complete subset of inputs that give a "yes/true" outcome ordinarily requires running an algorithm for each input. For Deutsch's formulated problem, the worse case scenario on a classical computer would involve running just more than half of the inputs. However, Deutsch formulated a quantum algorithm that would solve his problem in a single run.}. By the mid-to-late 90s, two prominent quantum algorithms had already sparked widespread interest by pointing out the relevance of quantum computing to real-world problems. Grover's algorithm demonstrated a polynomial improvement over classical methods for database search problems\cite{grover1996fast}, whereas Peter Shor theorized an approach for prime number factorization with a superpolynomial speedup over the best known classical algorithm\cite{shor1994algorithms}. The latter algorithm has critical implications on the future of cybersecurity as it would render certain public-key cryptographic schemes, like RSA, as insecure \cite{rivest1978method}.

%P3 - Quantum hardware & industrial growth - 341 words
As for executing quantum algorithms on a quantum physical platform, these were first reported independently by two research teams in 1998. Jones, Mosca and Hansen from Oxford, as well as Chuang, Gershenfeld and Kubinec in the US, reported using bulk nuclear magnetic resonance techniques on molecules for implementing Grover's search and Deutsch's algorithm \cite{jones1998grover, jones1998implementation, chuang1998experimental}. Deuterated cytosine and chloroform molecules containing  $^1$H and $^{13} $C nuclei acted as spin quantum bits (qubits) to perform the computations. Even so, the quantum circuit implementations were rather small-scaled, distinguishing only two qubits. The challenges of building practical and scalable quantum processors thus became apparent, leading physicist David DiVincenzo to contemplate the necessary criteria for a quantum computational platform\cite{divincenzo1995quantum, divincenzo1998quantum, divincenzo2000physical}. In essence, suitable platforms need to have a well-characterized quantum system which is  relatively robust to noise. The system must be reliably prepared in some initial reference state, and the platform must support a precise set of operations to manipulate the qubits' states according to an arbitrary quantum circuit. Qubit-to-qubit interactions need to be enabled  when the computation requires it, but the system should still maintain relatively long coherence times prior to measurement. An accurate readout of any one (or many) qubit state at the end of computation is also essential. To minimize noise and maximize qubit lifetimes, most quantum computing platforms operate under cryogenic conditions \cite{parker2022controlling}.  

Amid these engineering hurdles, major strides have been made in bringing practical quantum computing closer to reality on a variety of quantum architectures, such as superconducting qubits, trapped ions, photons and spin qubits \cite{majidy2024building, awschalom2025challenges}. In 2011, the company D-Wave was the first to market and sell a type of special-purpose quantum computer for solving optimization problems called a quantum annealer \cite{dwave2025}. In 2016, universal gate-based quantum computing became cloud-accessible with the launch of the IBM Quantum Experience \cite{ibmq2025, javadi2024quantum}. Currently, the number of operational quantum computers is estimated between 100 to 200 globally \cite{zeit2025, squant2025}. There are at least 178 universities, 451 private startups and 38 public companies that engage in quantum computing research and development \cite{QCR2025}. Multinational tech giants like IBM\cite{ibmq2025}, Google\cite{googleq2025}, Microsoft\cite{microsoftq2025}, Intel \cite{intelq2025} and Amazon\cite{ambrak2025} have invested heavily into expanding their quantum computing stack which spans both hardware devices and software tools.

 Quantum technologies also attract billions of dollars worth of investments by venture capital firms and government agencies\cite{QIW2025}. In 2025, global investments in quantum technology were evaluated at \$55.7 billion USD. According to Quantum Technology Monitor reports, recent technological breakthroughs in improved quantum hardware has pushed the projected market size of quantum computing to \$131 billion USD in the year 2040\cite{QTM2025}. By the same report, governments across the globe have committed \$1.8 billion USD to quantum technologies in 2024. Particularly for Canada, the National Quantum Strategy program was launched at the beginning of 2023 in order to position the country among early leaders in the industry \cite{canadaquantum2025}. The program aims to support research advancement, the development of a talented quantum workforce, and commercialization of quantum technology. One study commissioned by the National Research Council Canada (NRC) revealed that by 2045 the quantum industry in Canada will be worth \$139 billion CAD with net returns of \$42 billion CAD, as well as employ some 200,000 people. 
% from \$93 billion to \$ 131 billion

%140words
As quantum computing technologies continue to develop and attract more attention worldwide, there is a growing need to educate people on quantum science. First, this is about expanding quantum literacy and organizational readiness for the imminent integration of quantum-based solutions into many industries. Next, it is about assembling the future workforce of quantum scientists, engineers, software developers, applications experts, analysts, educators and consultants \cite{kaur2022defining}. The main challenge lies in demystifying a field that is known to be highly technical, interdisciplinary, and niche, even to university STEM students\cite{aiello2021achieving, greneirt2023future}. At many universities, courses dedicated to quantum physics are offered only towards the end of an undergraduate physics degree since knowledge of calculus and linear algebra is a pre-requisite\cite{stadermann2019analysis}. Courses dedicated to quantum computing are reserved for graduate-level studies or specialized programs as they require knowledge spanning physics, computer science, math and engineering \cite{haghparast2024innovative, goorney2025quantum}. Outside of structured academic programs, many excellent online resources are openly accessible to those wanting to learn and gain skills in quantum programming, algorithms and its applications through well-established platforms\cite{ibmqlearn2025, pennylane, micquant, quantcount, qosf}. However, these resources are likely intended for post-secondary learners or researchers in STEM. Overall, it is commonly assumed that learners have some working knowledge of linear algebra, scientific programming and quantum mechanics. 

Recently, it has become apparent that quantum topics should be introduced earlier in STEM education in order to make quantum career pathways accessible to a larger and more diverse subset of the population  \cite{nita2023challenge, He2021litreview, darienzo2024review, oikonomou2025enhancing, hasanovic2023quantum}. With this objective, valuable contributions have been made in developing quantum science workshops and learning materials designed for high school students\cite{angara2021teaching, tappert2019experience, walsh2021piloting, ivory2023quantum, satanassi2021quantum, satanassi2022designing, hughes2022teaching, perry2019quantum, sun2024computing, billig2018quantum, iqc_quantum_for_educators}. As might be expected, these vary considerably with respect to program content, experimental focus (e.g. optics, communications), program length (days, weeks or year-long), pedagogical methods (e.g. hands-on activities, experimental demos, games, computational exercises) and degree of mathematical rigour. The consensus is that teaching quantum physics concepts at the level of high school is a challenge in many forms and requires a rationalized approach strategy \cite{stadermann2019analysis, darienzo2024review, henriksen2014relativity, stokking2000predicting}. Effective teaching strategies often include: (1) the use of multiple learning modalities such as physical demos, hands-on exercises, group work and computational tools \cite{darienzo2024review}, (2) outlining relevant real-world applications \cite{henriksen2014relativity, hasanovic2023quantum, stokking2000predicting}, (3) applying analogies in describing unfamiliar concepts \cite{hasanovic2023quantum, He2021litreview}, (4) teaching the ``nature of science" by showing the historical progression of scientific thinking and developments in the methodology \cite{stadermann2019analysis}, and (5) focussing on qualitative over mathematical descriptions\cite{krijtenburg2017insights}.  

With these guidelines in mind, we have designed the Intro2QC quantum computing workshop that is described in the present paper. The paper is structured as follows. Sec. \ref{sec:goals} outlines briefly the goals and approach rationale of Intro2QC. Sec. \ref{sec:structure} summarizes the delivery structure of the workshop.  Sec. \ref{sec:resources} describes the resources we have developed and made available online. Sec. \ref{sec:content} recounts the workshop content and some supporting details as context for readers. Finally, Sec. \ref{sec:discussion} provides more in-depth discussion on the instructional design and approach rationale. 

\section{Pedagogical Goals and Approach Rationale}\label{sec:goals}

Intro2QC is a single-session outreach workshop dedicated to high school STEM students in Grades 9-12. The workshop was developed with the following intentions: (1) to rouse students' curiosity in quantum science and quantum computing, (2) to highlight the relevance of quantum computing in the modern world via real-world applications, (3) to introduce a conceptual basis for quantum physics by discussing properties like superposition, interference and entanglement,  and (4) to further illustrate quantum physics ideas with a practical tutorial in quantum programming. 

In formulating our approach for the workshop, the goal was not to bombard students with difficult math or overload them with tons of new information. Rather, students should have time to progressively assimilate these radically new concepts. We facilitate this incremental learning process by multiple means. Firstly, we sought to establish an interactive, judgement-free learning atmosphere that encouraged students to openly ask questions and give answers to the learning activities by emphasizing participation over correctness. Next, we wanted to apply a "motivation-first" logic for introducing quantum. In Part 1 of the workshop, instead of diving head-first into unfamiliar territory, we address the challenges of regular computing through a constructed analogy based on everyday life. Students relate easily to the example given and participate in mini exercises to discover the main difficulty in classical computation for certain types of problems. The mini in-class exercises in this section are also aided by a problem graphic, so simple logic would suffice to deduce the correct answers. By the end of this section, students are confident and already curious to learn how quantum computing can be useful in tackling the issues highlighted within real-world situations.

In the main section of the workshop, central ideas of quantum mechanics are introduced by walking students through the Schrödinger's cat gedanken experiment\cite{schrodinger1935present} and the Einstein-Podolsky-Rosen (EPR) paradox\cite{einstein1935can} with a plush toy demo. This teaching tool amplifies engagement as it is amuses students and makes the session more memorable overall. Reference is also made to the popular historical debates between scientists like Albert Einstein and Niels Bohr \cite{bohr1996discussion}, giving students a realistic perspective on the development and subsequent validation of quantum mechanics. Before the topic of quantum computing is explored, students are encouraged to share their thoughts on the potentially conflicting ideas within quantum physics. This triggers a short diversion onto the local realism assumption and the existence of alternative interpretations of quantum mechanics. Such topics are extremely effective in appealing to imagination and increasing excitement about quantum science. Next, quantum computing is broached by connecting how quantum properties might be exploited to address the computational challenges concluded in the first segment of the workshop.

In the final portion, students are guided through an online quantum programming tutorial on the Intro2QC website enabling them to rediscover quantum physics concepts in practice. Most of the tutorial code is already written on the website and only needs to be edited lightly on the students' end. The mini coding challenge questions provide an opportunity for them to draft their own code from scratch, but students may also choose to copy-paste some of the tutorial code as a starting template. The answers for all coding challenges are available on the website within hidden menus. The content of the first two sections of the workshop is also incorporated in the Intro2QC website. For this reason, students are invited to not take any notes during the workshop, and fully engage in the session. Enthusiastic students can also attempt the additional coding challenges listed on the website after the workshop.

The Intro2QC workshops were carried out during IYQ with about 160 high school students from schools within Greater Victoria in the Canadian province of British Columbia. Verbal feedback from both students and teachers indicated that the educational material and workshop delivery were engaging and appropriately matched to the students' level. We think that this workshop is a fitting introduction to quantum science and quantum computing for the average high school STEM student. There are no hard pre-requisites for students undertaking this workshop, although familiarity with terms like ``bits" and ``binary code" might be assumed for grades 9 - 12. Basic programming skills might prove beneficial but are not asked for.

In terms of the classroom set-up, access to a central projector and screen, or another central display, is necessary for the first two segments. Students also need access to contemporary computing devices and the internet for the programming portion. Desktop and laptop computers are ideally recommended, although most modern tablets and phones work as well. The presentation slides and full lesson plan have been added to an open-access online repository. The teaching plan should give STEM teachers enough confidence in exploring this topic with their students, even with little to no background or training in quantum science. The current paper might also provide educators with some additional background knowledge of the field at large.

\section{Workshop Structure}\label{sec:structure}
%252 words

The Intro2QC workshop comprises three (3) parts with a total time of 80 to 90 minutes. The first part highlights the challenges of computation for real-world problems with a suggested time of 20 to 25 minutes, including the mini exercises. The second part introduces key concepts in quantum mechanics and quantum computing and normally lasts 30 to 35 minutes to allow for some open discussion. The third section is an online tutorial in quantum programming on the Intro2QC website which can be completed in about 30 minutes. This might be less if students have some prior experience with computer programming. 

The workshop is delivered presentation-style for the first and second parts, with short intervals for interactive in-class exercises and discussion. For the third segment, students navigate to the workshop's website (\url{https://intro2qc.uvic.ca}) for the quantum programming tutorial. Only the \textit{Programming Exercises} page on the website is meant to be completed within the allotted time of the workshop. Additional coding activities are found on the \textit{Extras} page. At the end of the workshop, students are encouraged to complete any of the additional programming exercises, or recap on the workshop content for Parts 1 and 2 by exploring the relevant website sections on their own time.

The workshop may be also be shortened to a 50-minute session. The abridged version places emphasis on the quantum programming tutorial, with a brief introduction to quantum science and computing preceding this. The key points of the first two workshop segments are condensed into a concise 20-minute introduction.

\section{Educational Resources}\label{sec:resources}
\subsection*{Lesson Plan and Teaching Slides}
%156 words
The lesson plan and teaching slides for this workshop is openly accessible at the Github repository link listed under \textit{Workshop Materials} section. The lesson plan identifies 39 key learning points for a full-length workshop. There are 9 key learning points in Part 1, 12 in Part 2, and 18 in the programming section. In just under 25 pages, the lesson plan provides the full list of workshop objectives with detailed descriptions of each teaching point and how to perform the classroom demos. The Intro2QC website contains some excerpts from the lesson plan, but is shortened and made to be more reader-friendly for high school students. Teachers should aim to read the lesson plan in its entirety for an in-depth understanding of the material as well as the goals of the workshop. The lesson plan also highlights optional prompts for generating classroom discussion and some additional context on different topics that teachers might want to be aware of. 

\subsection*{Intro2QC Website}
%245 words
The Intro2QC workshop features an interactive website with executable code cells in the programming tutorial section generated with Jupyter book \cite{jupyterbook2020}. Jupyter book is an open-source tool for rendering static HTML pages from Jupyter notebooks and markdown files which are compiled into a ``book'' format with chapters and sections. The HTML outputs can be hosted on the internet using a web hosting service. Websites made with Jupyter book are commonly used as internet-based learning tools for delivering educational content where programming exercises form part of the curriculum. This type of interactive, browser-based computing experience eliminates the need for locally installing programming software, thereby reducing the overall complexity, hardware and time requirements of the Intro2QC workshop.

\begin{figure}[h]
\centering
\includegraphics[width=0.8\linewidth]{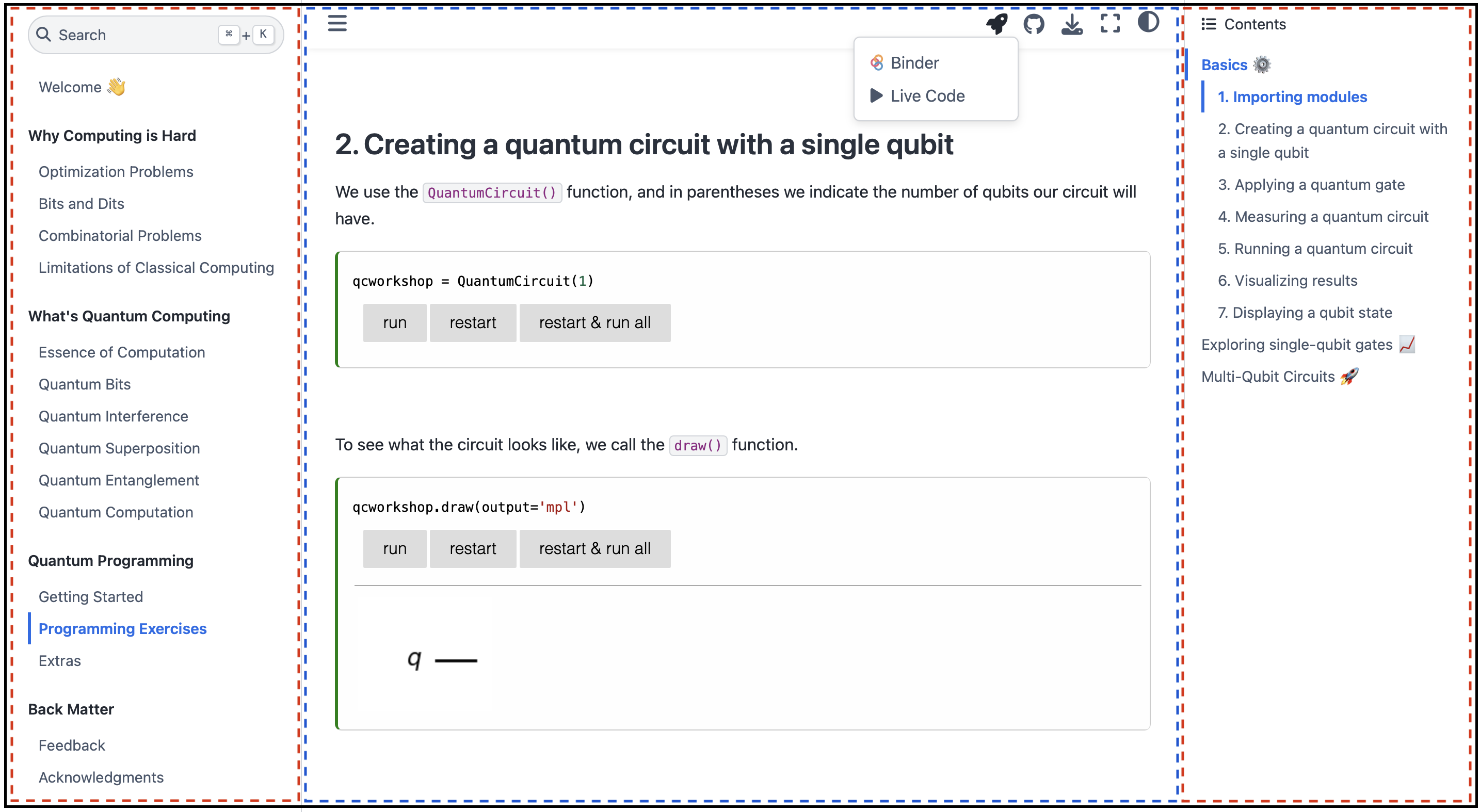}
\caption{Layout of the Intro2QC quantum programming tutorial. The Jupyter book website features left and right navigation bars (red dotted lines) and central page (blue dotted line).}
\label{fig:website}
\end{figure}

Figure \ref{fig:website} is a screenshot of the Intro2QC website highlighting layout and functionalities. The left and right sidebars are outlined in red, the main page and top menu in blue. The left sidebar indicates the current page and navigation links to different pages on the site. The right sidebar is useful for skipping to different sections and subsections on the current page. Code cells become executable by selecting the rocket icon on the top menu, followed by the \textit{Live Code} option. Once the kernel is ready, the code on the current page may be edited and run to obtain outputs. Also from the top menu, students can (1) access the site's Github repository, (2) download markdown and PDF files of the website pages, (3) enter into fullscreen mode and (4) toggle between light and dark modes of the website.

\section{Workshop Content}\label{sec:content}
\subsection*{Part 1: Why Computing is Hard}
% ~840 words 
The Intro2QC workshop aims to spark students' curiosity in quantum computing by illustrating its potential to more efficiently solve certain optimization tasks that are commonplace in the world \cite{abbas2024challenges}. Combinatorial optimization problems readily illustrate the challenges of classical computing and are likely the most impactful applications of quantum computing in the near-term. Prominent examples are the travelling salesman problem (TSP), the Max-Cut problem, the knapsack problem, bin packing, graph colouring and job-shop scheduling to name a few \cite{lucas2014ising}. These are all are classed as NP-hard problems\footnote[2]{Computational decision problems can be classed according to how difficult it is to find a solution and verify its correctness. P (polynomial time) problems are computationally efficient to solve and verify. NP problems (non-deterministic polynomial time) are those whose solutions can be verified in polynomial time, but not necessarily determined as easily. NP-hard problems are presumed the hardest problems in the NP class. An algorithm that can find the solution to an NP-hard problem in polynomial time would also solve all other NP-problems in polynomial time.} which become computationally intractable for a large number of decision variables. Optimization problems of this type are prevalent across many sectors and industries, including finance, manufacturing, scheduling and logistics, telecommunications, electrical power distribution, healthcare, drug discovery, and materials design \cite{herman2023quantum, kurowski2023application, phillipson2023quantum, zhou2022quantum,  flother2023state, bauer2020quantum}. 

While unlikely that computational decision problems are formally treated in the classroom at this level, high school students are already confronted with the difficulties of decision-making in their everyday life.  In teaching basic combinatorics, the universal model is the ice cream shop analogy. In an ice cream shop, a customer decides how they would like their ice cream prepared. First is the choice between a cup and a cone. Next is selecting the flavour of the ice cream, and finally the topping to complete the order. Given the sequential ordering for each new decision and the finite number of options at each step, one can directly apply the rule of product or the fundamental counting principle in combinatorics \cite{wallis2016introduction, vasquez2021teaching}. 

\begin{figure}[h!]
\centering
\includegraphics[width=0.9\textwidth]{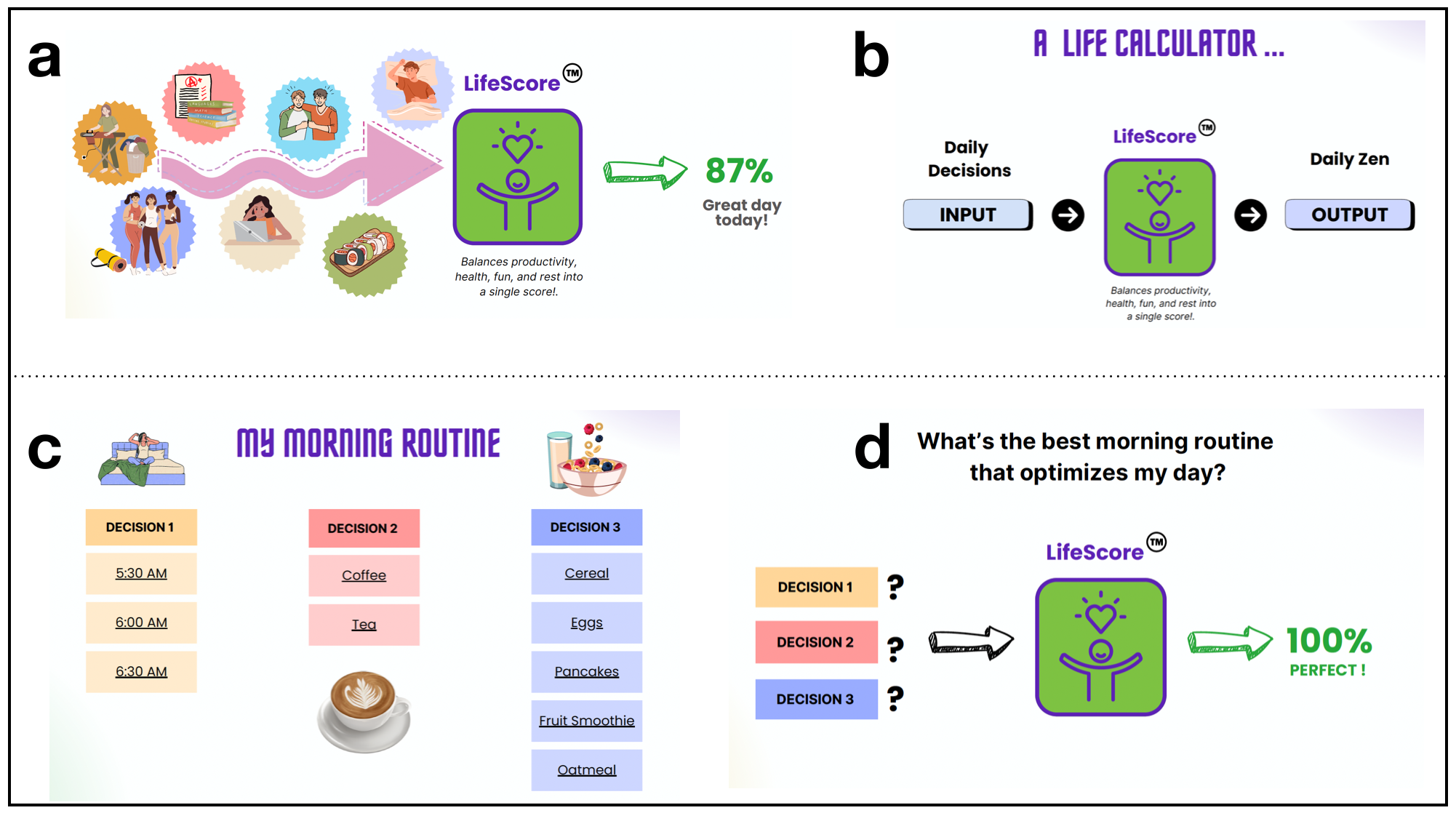}
\caption{A daily life analogy of an optimization problem. \textbf{(a-b)}. The operation of the special-purpose calculator, LifeScore, a blackbox function, which scores a set of daily decisions. \textbf{(c-d}) The optimization problem is a simplified scenario of $3$ decision variables and a set of degrees of freedom for each variable.}
\label{fig:lifescore}
\end{figure}

In Intro2QC, we construct a similar analogy on decision-making in an adolescent's life that creates an impactful introduction to the workshop. We pose that there exists a special type of calculator that computes a score based on a student's morning routine, and that the routine that maximizes this calculator's output must to be determined. It is assumed that there is at least one optimal answer which gives the perfect score. The graphics in Figure \ref{fig:lifescore} \textbf{(a-b)} illustrates the functioning of the calculator, whereas Figure \ref{fig:lifescore} \textbf{(c-d)} poses the problem at hand. In more scientific terms, the special purpose calculator represents some complex, multivariable function or even a blackbox function, where finding optimal solutions analytically is assumed to be difficult or impossible. In this scenario, an exhaustive search of all possible inputs and their computed output scores is required to find the solution(s). Students are first asked to deduce the total number of possible morning routines for the problem, which can be determined graphically with the aid of Figure \ref{fig:lifescore} \textbf{(c)}. 

The rule of product formula may also be verified via this method. The simplified mathematical statement for this counting principle is the following. For $N$ decision variables, each with its own degree of freedom in $D = \{d_1, d_2, \dots, d_N\}$, the total number of combinations $C$ is simply: 

\begin{equation}
C = d_1\times d_2 \times \dots \times d_N
\end{equation}

In the present scenario, there are $N=3$ decision variables for which $d_1 = 3$ degrees of freedom for the first variable, $d_2 = 2$ degrees of freedom for the second variable, and $d_3 = 5$ degrees of freedom for the last. This gives $C = 30$ possible morning routines, all of which must be run through the calculator to resolve the optimal input. Naturally, students can deduce from the rule of product that if either the degrees of freedom of any variable in $D$, or the number of variables $N$ are increased, then the total possible routines $C$ climbs rapidly as these numbers must be multiplied together.  For variables with each the same degree of freedom $d$, i.e. $d_1= d_2 = \dots = d_N = d $, this is a clear illustration of an exponentially growing function since $ C= d^N$. 

To assist with computational thinking, the constructed analogy is reframed to a  binary optimization problem. Since computers process information in binary, computational input is fundamentally encoded as bitstrings of a specified length. The British Columbia K-12 curriculum covers the binary system in Grade 11 Computer Science, and even earlier in Grade 8 under Applied Design, Skills, and Technologies (ADST) \cite{bccurriculum1, bccurriculum2}.  Students are therefore expected to be moderately familiar with terminology like "binary" or "binary code" at the high school level. 

For a binary variable problem, we consider that the input to our special purpose calculator is a bitstring consisting of $N$ bits. Each position in the string represents a separate decision variable, and the bit value at that position (either `$0$'or `$1$') indexes one of the two options available for that decision variable. With the rule of product formula, students conclude that there are $2^N$ distinct bitstrings that must be passed as input to the calculator in order to find the optimal string that maximizes the output score. Thus, exhaustive search for binary variable problems scale exponentially as $2^N$ for $N$ variables each with 2 degrees of freedom. 

With access to one calculator which takes only $N$ bits of input at once, the complexity of the computation is exponential in time since $2^N$ successive computations must be performed. On the other hand if we have access to $2^N$ calculators to run all bitstrings simultaneously, the computational complexity is exponential in its memory requirements. It is important to inject these ideas on parallel computation to establish a foundation for the quantum computing discussion in the upcoming section. 

\subsection*{Part 2: What's Quantum Computing?}
\begin{figure}[h]
\centering
\includegraphics[width=0.8\linewidth]{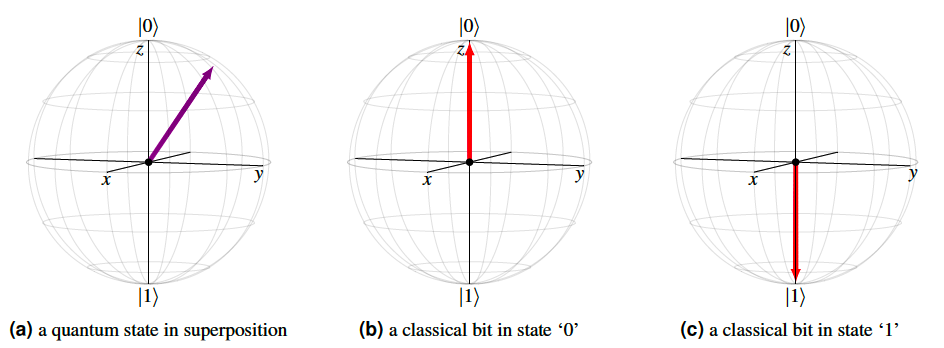}
\caption{The Bloch sphere for visualizing quantum states. Classical bit states are depicted in red and the quantum bit state in purple. The quantum state indicated here has a higher probability of being measured as `0'.}
\label{fig:bloch_tikz}
\end{figure}

%~1621 words without equations, footnotes, boxes
As the students are now reacquainted with the concept of computational bits and bitstrings, it is beneficial to introduce the quantum bit or qubit. A quantum bit is the basic unit of quantum information. As depicted in Figure \ref{fig:bloch_tikz} \textbf{(a)}, the state of a quantum bit can be illustrated as an arrow vector in a unit sphere called the Bloch sphere. The arrow tail is initiated at the centre of the sphere and the arrow tip touches the sphere's surface. In Figure \ref{fig:bloch_tikz} \textbf{(b)} and \ref{fig:bloch_tikz} \textbf{(c)},  the state of a classical bit is drawn as an arrow pointing upward towards the north pole if it is in the state `0', or downward towards the south pole if it is in the state `1'. Therefore a classical bit is only allowed to point in one of two directions in the Bloch sphere representation. On the other hand, the quantum bit is permitted to point toward any direction on the sphere's surface including upward and downward. 
 
In more technical terms, the allowable states for classical bit is a discrete set with only two possible values, i.e $ |\{0, 1\}| = 2$. The mathematical space of this set is said to be 1-dimensional since its Bloch vector can only point along the $z$-axis of the sphere. The allowable states for a quantum bit is an infinite set of vectors that are representable on the 3-dimensional Bloch sphere \footnote[3]{Formally, the dimensionality of a qubit is the 2-dimensional complex Hilbert space. There are two complex numbers, and hence four real numbers required to describe the state of a qubit. Due to the normalization condition and considering only relative phase (and not global phase) of the two complex numbers, there are actually two degrees of freedom for a pure single qubit state. In relation to the Bloch sphere representation, the degrees of freedom are the two angles used to describe the Bloch unit vector which are the polar $\theta$ and azumithal $\phi$ angles. }\footnote[4]{Pure single-qubit states are said to be 2-dimensional as they are represented by vectors that point only to the 2-dimensional surface of the 3-dimensional Bloch sphere. Mixed single-qubit states are those that can be found within the interior of the Bloch sphere. These are 3 dimensional since the radial component of the vector must now be defined to include vectors not ending at the surface. In quantum information, a mixed state is the result of a probabilistic combination of pure quantum states. It usually denotes that there is some unknown information on the state of the qubit. As a result, the Bloch sphere depiction is most appropriate for describing pure single-qubit quantum states, rather than mixed or entangled qubit states.}. 

\subsubsection*{Properties of Quantum Mechanics}

The Bloch sphere picture of the allowed states for a classical bit versus a quantum bit is sufficient to initiate the discussion on key properties and behaviours of quantum objects. In the Intro2QC workshop, we focus on superposition, state collapse (or wavefunction collapse), interference, and entanglement. From the Bloch sphere representation in Figure \ref{fig:bloch_tikz}, the most apparent one is superposition which is the ability of a quantum object to be in multiple classically-observable states at the same time. The observable states refers to the states `0' and  `1' . Hence, a quantum bit can have a state that is simultaneously in some proportion of state `0'  and  state  `1' .\footnote[5]{The quantum state examples used in this workshop are restricted to two-level systems, in reference to binary computation with only 2 degrees of freedom. While systems with higher degrees of freedom are far more common in nature, they are harder to control for quantum computation. If time permits, students may like to know that a \textit{qudit} is the term used to refer to a unit of quantum information for a system with $d$ degrees of freedom when  $d$ is larger than 2}. Whenever the Bloch vector is not pointing exactly upward or downward, the qubit is in a state of superposition. However superposition is only possible when a qubit has not yet been measured (observed). When this happens, the superposition state of the qubit collapses probabilistically to one of the classically observable states. That is to say that a qubit inherently loses its \textit{quantumness} once measured.

To gain knowledge of the state prior to measurement, one must repeatedly prepare the qubit in the same way and record the measurements to determine the outcome fractions for states `0' and `1'. These probabilities are a partial description for the quantum state.The mathematical description of a quantum state is similar to a probability-weighted average of each classical state. In pseudo-math, this can be written as follows: 
\begin{equation}
\text{Qubit state} =  c_0 \times \text{state `0'}  + c_1 \times \text{state `1'}
\label{eq:qstate1}
\end{equation}
where $c_0$ and $c_1$ are coefficients relating to the probabilities of each outcome respectively \footnote[6]{These coefficients are referred to as \textit{probability amplitudes} rather than probabilities. In actuality, they are complex numbers whose squared modulus give probabilities which sum to 1:  $|c_0|^2 + |c_1|^2 = 1$. The correct mathematical expression for the general qubit state differs from the one given to students only by the use of Dirac bra-ket notation and specification of the normalization condition for $c_0$ and $c_1$.}. These describe the amount of the qubit state that is in the `0' state and in the `1' state.

The quantum superposition principle is encapsulated into a simple teaching demo for students. Box A outlines Schrödinger's popular thought experiment on the puzzling nature of superposition in quantum mechanics.  

\begin{tcolorbox}[title= Box A - Schrödinger's Cat Teaching Demo]
\begin{enumerate}
\item For this demo, we use a Schrödinger's cat plush toy which depicts a cat that is in an `alive' state on one side and a `dead' state on the opposite side. 
\item We pose that the cat is a quantum object and in a state of superposition. While not looking at the cat, we use our hand to move the cat in such a way that it switches between the `alive' and `dead' state many times to indicate superposition. 
\item We have decided to observe the state of our quantum cat. Once we decide to look at our cat, we stop moving the cat so that only one side is shown to the students. We repeat the demo for the students showing random outcomes each time and ask the students to describe what happens once the cat is observed. 
\item After learning about the collapse of a quantum state under observation, we then ask students how might they be able to learn the cat's quantum state prior to collapse. We hint that we are able to prepare the same quantum cat as many times as we wish.
\end{enumerate}\end{tcolorbox}

The state of the cat in the demo outlined in Box A is also expressible as in eq. \ref{eq:qstate1}. Here, the `0' state can be the `dead' state of the cat whereas the `1' state reflects the cat's `alive' state. Additionally, if both states of the cat are equally probable, then the weighting (norm) of the numbers $c_0$ and $c_1$ would be equal. 

Another central quantum property that strongly confirms the wave-like nature of quantum systems is interference. From classical physics, waves have long been known to interact in intriguing ways. Waves can: reflect off of boundaries, bend (refract) when moving from one medium to another, and spread out from its source, or a hole or gap encountered at a barrier. Wave interference occurs when multiple spreading waves of the same kind become superposed in a stable manner. When the amplitude of each wave is adds up to a greater amplitude, then interference is said to be constructive. Oppositely, if the resultant wave has lower amplitude than the original waves, the interference is destructive. Perfect everyday examples that are relatable to students are water ripples on the surface of a body of water and noise-cancellation technology. Similar to classical waves, quantum objects are able to interfere in this manner. The notion that quantum objects exhibit both wave-like and particle-like behaviours is termed ``wave-particle duality".

Finally, entanglement is a strange phenomenon that can arise when two or more quantum objects interact in a certain manner. It results correlations in measurements for each of the entangled objects which cannot be rationalized from a classical physics perspective. In an unentangled quantum system, the mathematical expression for the state of the system can be factorized into individual expressions for each quantum object making up the system. In contrast, the mathematical expression for an entangled state is not separable in this manner. The odd implications of quantum entanglement was historically highlighted in the EPR paradox. A modification of the first teaching demo on the Schrödinger's cat thought experiment is applied here to illustrate the idea of entanglement to students. This demo is summarized in Box B. 

\begin{tcolorbox}[title= Box B - EPR Paradox Teaching Demo]
\begin{enumerate}
\item Students are shown two Schrödinger cat plush toys with their `alive' and `dead' states.
\item  The cats are in superposition (not yet measured) and have interacted in some manner at a point in time. This is indicated by bringing the toys together, but rotating them between the `alive' and `dead' states many times.
\item Separating the cats at arm's length, we explain that after the initial interaction they have been separated by a very large distance. One cat is placed on one end of our solar system and the other at the opposite end.  
\item Two observers are now placed to evaluate the cats' states at each location where they will conduct their measurements simultaneously. Upon measurement, both cats have now collapsed to the same state. 
\item  Posing that the entangled cats are prepared in the same way, we repeat the demo giving a different correlated collapse. For example, both `alive' in the first trial then both `dead' in the second. After multiple repetitions, students deduce that measurements of both cats give the same result within each trial.
\item Students are asked to reflect on what it might mean if the results are always linked despite being measured at the same time. Here we should re-emphasize that the quantum state only collapses at the time of observation, and the states are not predetermined. 
\end{enumerate}\end{tcolorbox}

From the scenario outlined in Box B, the entangled state represented in qubits is generalized to the following expression, where the state `00' symbolize both cats found in the `dead' state and the `11' where both cats are `alive'. 

\begin{equation}
\text{Entangled state} = c_{00} \times \text{state `00'} + c_{11} \times \text{state `11'}
\end{equation}

In the above expression, there is no way to factorize the terms to independently describe each qubit in the system. Obtaining state `0' for the first qubit constrains the second qubit also to `0', and similarly obtaining the outcome `1' for the first qubit constrains the second to `1'. A completely separable system would mean that there is a state for each qubit irrespective of the what the other is. Then, if the first qubit is in `0' the other could be in `1' without limitations (and vice versa), contributing to non-zero probabilities for measuring the combined state  `01' and `10' for the system.
 
The surprising conclusions of quantum mechanics, as highlighted by these thought experiments, STEM students deeper discussion of the reality of nature which may be explored with students if time permits. Albert Einstein and collaborators\cite{einstein1935can} viewed quantum entanglement as a violation of the theory of special relativity which asserts that the speed of light is the universal speed limit. No type of matter, energy or information can travel faster than this constant. Correlations in measurement outcomes of this kind would insinuate that information is passed between the entangled objects at the exact moment of collapse, which makes it faster than the universal speed limit.  Rather than "spooky action at a distance", Einstein concluded that quantum mechanics was not a complete description of nature, and that there were hidden variables not yet accounted for in quantum theory that determine the eventual measurement outcomes. However, Einstein's viewpoint has been experimentally demonstrated as invalid by Nobel Prize winning experiments on entangled light conducted between the early 70s and late 90s \cite{clauser1969proposed, freedman1972experimental, aspect1982experimental, bouwmeester1997experimental, pan1998experimental}\footnote[7]{Some 29 years later, John S. Bell mathematically proved that hidden-variable theories would have a lower level of measurement correlation than what is predicted by quantum mechanics. In the following decades, photon experiments demonstrating the violation of the correlation limit set by Bell would corroborate the veracity of quantum mechanics, and reject hidden-variable models. The 2022 Nobel Prize in Physics was awarded to John Clauser, Alain Aspect and Anton Zeilinger for conducting experiments with entangled photons and testing Bell's inequalities}. Presently, quantum mechanics is understood not to violate Einstein's universal speed limit. It rather rejects the idea that information transfer between entangled objects is needed at all. The realization of quantum entanglement experiments over decades seemingly reveal that the universe is not locally real.\footnote[8]{The core of Einstein's disagreement with quantum theory springs from an assumption of ``local realism". \textit{Realism} reflects that objects have defined properties whether we measure them or not. \textit{Locality} is when these properties are influenced only by near neighbours in their environment. Quantum mechanics challenges the local realism assumption as shown by the correlated measurement outcomes of space separated entangled quantum objects.}

\subsubsection*{The Essence of Quantum Computation}

By exploiting the counterintuitive quantum properties discussed previously, quantum computing intends to solve particularly challenging problems in a more efficient manner than traditional computation allows. The latter hinges on manipulating the states of classical information (bits) with classical logic operations, where quantum computation entails operating on quantum information (qubits) with quantum logic operations. Reconsidering the challenges in computing for optimization problems pointed out in Intro2QC Part 1, an appropriate justification for quantum computing might be delivered to students in the following way. If solving an optimization problem for $N$ binary decision variables requires an exhaustive search on $2^N$ bitstring inputs, it is reasonable that $N$ qubits in superposition is simultaneously representative of all $2^N$ inputs. Quantum superposition thus permits encoding large-sized problems on exponentially-less computational memory! 

Despite the last point, we cannot however retrieve all $2^N$ outputs in one run due to the collapse of the superposition state upon measurement. Hence quantum interference and entanglement become crucial ingredients to achieving an advantage via quantum computation. For a given problem, constructive quantum interference can amplify the outcome probabilities associated with optimal answer(s) while destructive interference suppresses all non-useful outputs. Furthermore, entanglement creates patterns across the whole system by encoding special correlations between qubits that are not efficiently represented classically\footnote[9]{Certain types of entanglement involving so-called Clifford gate operations are easily simulated on classical hardware (Gottesman-Knill theorem). Entanglement generated with non-Clifford gates are hard to simulate classically.}. These correlations enable quantum interference in a highly coordinated way.  Superposition, entanglement and interference are altogether elemental to the design of powerful quantum algorithms. 

\subsubsection*{Quantum Computers}

A computer is a programmable device that can process computational information according to some set of instructions and produce relevant output to be interpreted by the user. At a fundamental level, computational information describes the states of the main physical component which make up computer chips: transistors. The voltage level in these transistors determine the value of the bits. If the voltage level is below a designated threshold value, the state of the transistor is read as `0'. Above that threshold voltage, it is read out as `1'. Logic gate operations alter the states of bits in keeping with programming instructions. After processing, the final states of the bits form the output.

Analogously, quantum processors manipulate the states of quantum objects in a quantum mechanical way before the final measurement. Quantum computing hardware varies by the type of qubit\footnote[10]{The different types of qubits include cold atoms, trapped ions, photons (particles of light), and superconducting circuits which are likened to artificial atoms.} and how qubit states are altered and read out. Qubits are kept away from external noise in well-controlled environments due to the sensitive nature of quantum states upon interaction and measurement. With the exception of photons, these environments are typically refrigerated to near absolute zero, which is colder than outer space. To perform a computation, quantum circuit instructions dictate the precise control of qubits using electromagnetic fields for creating superposition and driving entanglement between qubits. 

Despite substantial advancements over the past decade, the current era of quantum technology is still relatively small-scaled and prone to noise. The field, however, is rapidly-growing and extensive research efforts seek to develop hardware capable of solving real-world applications in the next decade or two. In the meantime, small quantum processors and classical simulations cover the bulk of the workload in quantum algorithms and applications research. Simulation involves the use of classical hardware or methods to emulate quantum mechanical systems and quantum computing circuits. The final segment of the workshop is an online programming tutorial on the Intro2QC website utilizing Qiskit for quantum circuit simulation. 

\begin{figure}[h]
\centering
\includegraphics[width=0.8\linewidth]{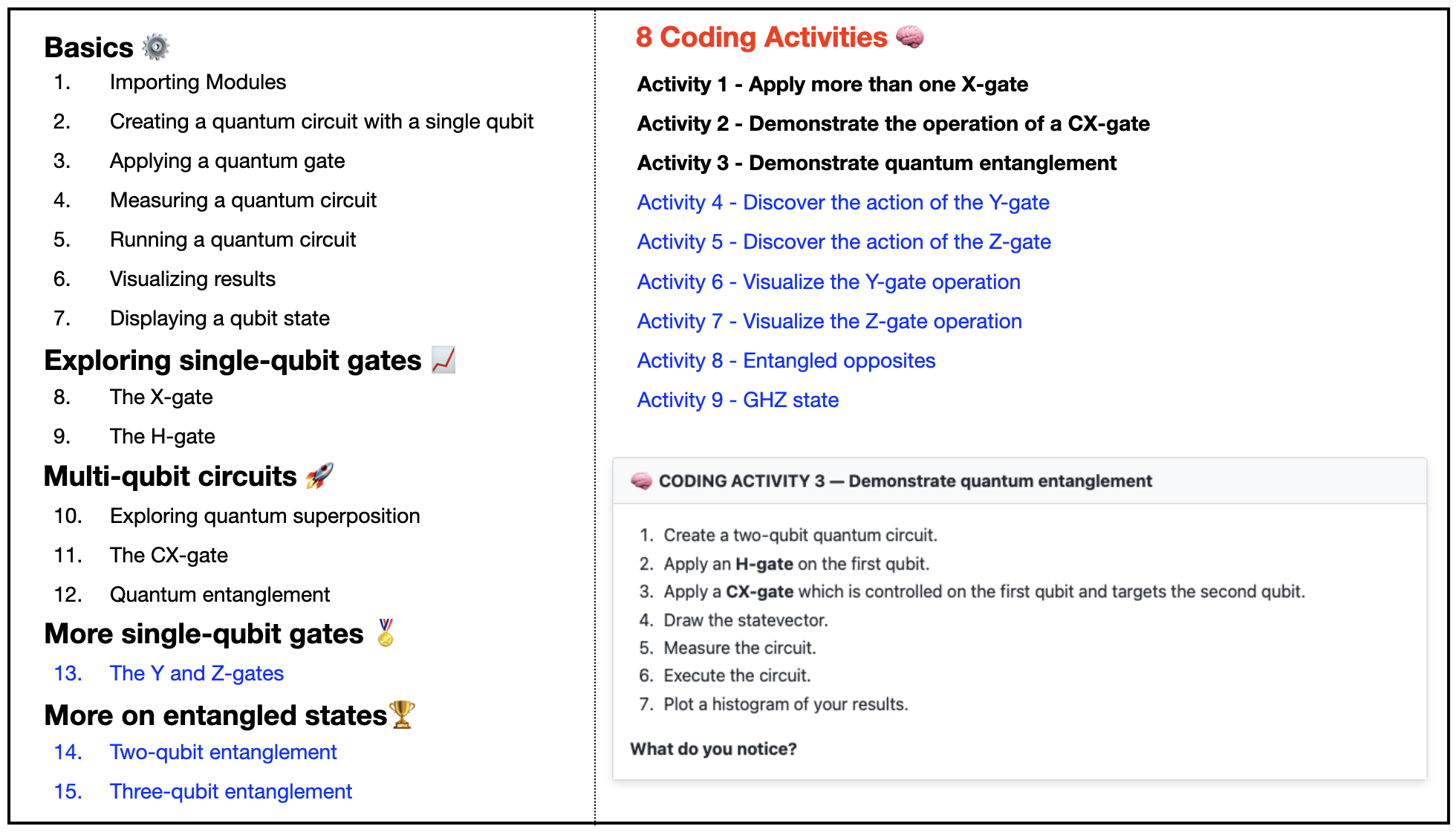}
\caption{Outline of the Quantum Programming section with coding activities. Objectives and activities in blue may be completed by students after the workshop.}
\label{fig:progout}
\end{figure}

\subsection*{Part 3: Quantum Programming}
% ~600 words 

Qiskit is an open source Python-based software development kit (SDK) for quantum programming developed by IBM, and is regarded as the most popular quantum programming language \cite{javadi2024quantum}. The IBM Qiskit SDK allows users to design quantum circuits by defining quantum operations on a set of qubits and to execute these circuits on a backend device such as real quantum hardware or classical simulators. In Intro2QC Part 3, the aim is: (1) to give students a tangible perspective on what quantum computing is like, (2) to help cement the crucial ideas covered in the Intro2QC Part 1 and Part 2, and (3) to build confidence in the knowledge gained by putting everything into practice.

Students log onto the workshop website, launch the interactive programming interface and follow along with the tutorial. All code is already visible for each tutorial block, except for coding activities. In the tutorial, students are asked to rename a few circuit variables as per their preference, which helps them to better understand what each line of code is doing as they keep track of the relevant circuit objects and attributes. Students are therefore engaged throughout the tutorial even though much of the code is provided. For a tutorial in programming under timing constraints, this is a time-efficient method suited to high schoolers with little to no programming experience. A larger portion of time is allotted for attempting the coding activities where they may copy tutorial code from previous blocks and edit it accordingly. 

Figure \ref{fig:progout} lists all the tutorial sections, blocks and coding activities. Those highlighted in blue are not counted in the allocated time, but are additional blocks and exercises that may be optionally completed by students. Only 3 of 9 coding activities form the core of the workshop, so students have ample time to write their own short lines of code without copy-pasting. Answers to all coding activities including the extra ones, are provided in the clickable drop-down tabs on the webpage. 

\begin{figure}[h]
\centering
\includegraphics[width=0.85\linewidth]{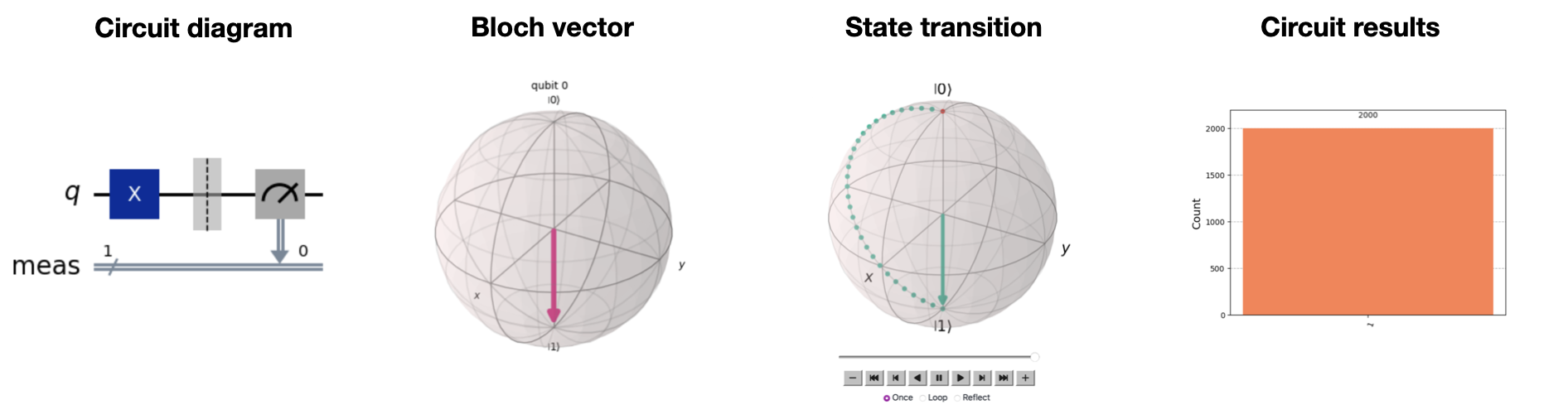}
\caption{Qubit state and circuit visualization options available in Qiskit.}
\label{fig:visuals}
\end{figure}

The programming tutorial fundamentally covers: how to start using Qiskit, the functions used to create quantum circuits, how to apply quantum gates, visualize qubit states, execute quantum circuits and obtain measurement outcomes. The continuous use of visualizations throughout the tutorial is an effective method for sustaining the students' engagement. It is also instrumental for boosting a conceptual understanding of quantum mechanics and computing. Figure \ref{fig:visuals} illustrates several of the visualization options for representing qubit states, circuits and circuit results in Qiskit using matplotlib \cite{hunter2007matplotlib}. Circuit diagrams give a snapshot of the current state of the quantum circuit object with all its qubits and applied gates. Bloch vector plots render the state of a single qubit on the Bloch sphere. State transitions animate the evolution of a single qubit state as gates are applied to the qubit. Histograms summarize the collective outcomes of the executed circuit over a specified number of shots. This emphasizes the probabilistic nature of measurements in the context of quantum mechanics. Students use these visualization options in the tutorial to self-check and debug their code appropriately during the coding activities. 

With programming basics covered, demonstrations of superposition and entanglement in action are facilitated by knowledge of just 3 quantum gates and how they each affect a qubit's state. These are the Pauli X-gate, the Hadamard gate and the controlled X gate. The Pauli X gate transitions a qubit in state `0' (north pole) to state `1' (south pole) and inversely. It is the equivalent action of a bit flip or the NOT logic gate in classical computing. On the 3D Bloch sphere, the qubit vector is rotated by 180$^\circ$ around the $x$-axis of the sphere. The second single-qubit gate is the Hadamard operation which creates an equal superposition state when a qubit is initially in state `0' (or `1'). The operation rotates the Bloch vector from the north pole (or south) to lie along the positive x-axis (or negative x-axis) in the equatorial plane. By setting up a $N$ qubit circuit and implementing Hadamard gates on each qubit, students notice that the measurement outputs form a fairly uniform distribution of all $2^N$ combinations of classical bitstrings. The controlled X gate is the only multi-qubit gate treated in the workshop, and the conventional example used to produce entanglement. Its operation is analogous the CNOT logic gate with a control bit and target bit. Only when the control qubit is in a state `1', a Pauli X operation is acted on the target qubit. Students experiment and decipher the operation of these quantum gates,  and use them to address the entanglement challenge exercise for creating a Bell state. In the optional programming content, more single qubit gates like the Pauli-Y and Pauli-Z gates are reviewed and students are able to construct slightly more complicated entangled states like the 3-qubit GHZ state. 

\section{Discussion} \label{sec:discussion}

Various reports of quantum outreach programs and STEM teaching experiences have informed and validated the scope, content and delivery of the Intro2QC workshop for achieving the maximum impact.  A literature review of quantum information science and technology (QIST) outreach programs for high schools conducted by Darienzo and Kelly \cite{darienzo2024review} outlined common teaching methods and best practices based on 10 workshops in recent years. Mainly, all workshops integrated several types of learning modalities, from hands-on activities with physical objects to pen and paper exercises, game-based learning, computational simulation and/or programming, group discussions and science experiments \cite{angara2021teaching, satanassi2021quantum, walsh2021piloting}. Multimodal learning offers multiple access points for students with diverse learning styles to actively engage in understanding the material \cite{bouchey2021multimodal}. In a one (or two) day workshop, Angara et al.\cite{angara2021teaching} used a doughnut pillow for demonstrating quantum concepts (one side with sprinkles and the other without), practice sheet exercises, a quantum-themed board game and quantum programming activities. Another example, Satanassi et al.\cite{satanassi2021quantum} blended traditional-style lectures, group-based learning for solving problem exercises and held group-focussed and collective discussions for exploring content topics over 6 workshop sessions of 3 hours each.  Overall, effectively merging multiple modalities enhances student interest, participation, comprehension and knowledge retention. The Intro2QC workshop applies multimodal strategies via physical demos with the Schrödinger cat stuffed toys, mental and on-paper short exercises, open class discussions, and computer programming with graphical simulations of quantum states and their evolution. 

Regarding the instructional design and arrangement of learning key points, our "motivation-first" method utilizing a constructed analogy of optimizing a personal routine with generalization to real-world applications, is an obvious extension of context-based learning. The latter works by adding recognizable elements into the learning process which augments students' perceived relevance of the topic and in turn motivates curiosity for learning the new material \cite{glynn2005contextual, ultay2014context}. In a review of the available literature on quantum computing efforts in K-12, He et al. \cite{He2021litreview} advocated "making quantum concepts relevant to everyday events". The authors made the suggestion based on the report by Moraga-Caldéron et al.\cite{moraga2020relevance} that upper secondary school students found that learning quantum was not personally relevant to them. Additionally, from experiences in developing a post-secondary quantum technician training curriculum, Hasanovic 2023\cite{hasanovic2023quantum} suggested that raising awareness of quantum science in a general audience should commence with real-world examples and applications before the introduction of quantum mechanics concepts .

Another strategy undertaken in Intro2QC is teaching "nature of science" (NOS)\cite{lederman2013nature}. Historically, it became evident that in order to "humanise" science, boost scientific literacy and expand the workforce, science education needed to diverge from the teaching of need-to-know facts and thus incorporate aspects of the process and epistemology of scientific inquiry\cite{jenkins2013nature}. NOS characteristics involve differentiating between scientific laws and theories,  observations and inferences, and an understanding that although science is empirically based, it involves creativity, is partially subjective, culturally embedded, and subject to change upon discovery of new evidence \cite{lederman2004revising, lederman2013nature}. Within this context, quantum science offers many opportunities for NOS teaching. Stadermann et al.\cite{stadermann2019analysis} compared curricula for upper secondary schools in 15 countries including parts of Europe, the UK, Australia and Canada, and analyzed the application of NOS in secondary level quantum physics. They listed specific quantum topics that accomplish NOS objectives. The evolution of physical theories of light from corpuscules to waves, then quanta with the emergence of wave-particle duality, are excellent for demonstrating the tentativeness of science. The thought experiments for Schrödinger's cat and the EPR paradox serve to highlight both scientific methodology and creativity in the field. The paradigm shift from classical perspectives on the nature of reality and the resultant philosophical implications of quantum theory were underlined in the controversial Bohr-Einstein debates, showing that scientific evidence can still be open to interpretation and even great scientists make assumptions in their deductions. Engaging students with these types of discussions helps establish a more reliable and approachable perspective of the field and the methods by which new knowledge is acquired. 

A final point is the intentional minimization of mathematical formalism in this workshop. In an extensive review of studies on introductory quantum science curricula at the secondary and lower undergraduate level, Krijtenburg et al. \cite{krijtenburg2017insights} concluded that a non-mathematical concept-focussed approach is a viable teaching strategy which leads students to an adequate understanding of quantum concepts. Dür and Heusler \cite{dur2016qubit} also remark that the qubit as a two-level system is sufficient for introducing quantum physics at high school level, including visualizations on the Bloch sphere, without need for mathematical formalism.  Recent publications \cite{rexigel2025investigating, bley2026modelling } have emphasized the importance of multiple external representations (MERs) within the DeFT framework (Designs, Functions, Tasks)\cite{ainsworth2006deft} for quantum science curriculum development. The number, sequence and translations between different representations (e.g. oral, text, diagrams, graphics, tables, equations, simulations) must be effectively chosen to support cognitive learning processes. According to the DeFT framework, MERs serve to provide complementary information, to constrain interpretation, and to construct deeper understanding of a concept. Specific to quantum science, some examples of multiple representations include 2D and 3D Bloch spheres, Q-spheres, quantum circuit diagrams, state vector equations, matrix/vector notation and Dirac notation. Here, we've considered the necessary set of representations for meeting the goals of the workshop within the short timeframe. We slowly incorporate and link together these representations as it fits the objective of each section. In Part 1, a graphic table aids students in deducing the total number of combinations in the routine optimization problem prior to introducing the combinations formula. In order to illustrate combinatorial explosion, there is a class activity in bit counting which employs a graphic with a horizontal line for each bit variable, in partial resemblance to quantum circuits to be introduced in Part 3. In Parts 2 and 3, the Bloch sphere picture is complementary to the mathematical state-vector equation for representing probabilities of measurement outcomes. It again comes in handy for demonstrating the actions of single-qubit gates on the qubit state as vector rotations on the Bloch sphere. The careful selection of graphical and simplified mathematical representations in Intro2QC is what enables students to quickly grasp and link together complex ideas in little time. 

Intro2QC workshop sessions were exceptionally well received as validated through informal feedback forms (optional and anonymous). A large majority of responders rated the workshop experience as 8 or higher out of 10. The workshop was also relatively easy to follow with almost all responses as 3 or higher out of 5. Responders also rated the effectiveness of the visuals as learning aids very highly. 

\section{Conclusion}
To summarize, this paper describes an approachable introductory workshop for quantum computing targeting grades 9-12 students pursuing STEM aligned curricula. Intro2QC is an exciting 90-minute session covering some core understandings in quantum mechanics, quantum computation and quantum programming without being overly complex or mathematical. We think that this resource is not only approachable to STEM students, but also STEM teachers who would like to explore quantum mechanics and quantum computing in the classroom. STEM awareness campaigns and outreach programs, like the Intro2QC workshop, are a crucial part of widespread global efforts meant to attract, develop and diversify the future generation of talent in the quantum industry.

\section*{Workshop Materials}
The presentation slides and lesson plan are available at \url{https://github.com/Jaim-gem/intro2qcworkshop} and the workshop website is hosted at \url{https://intro2qc.uvic.ca}.

\section*{Acknowledgements}
The Intro2QC project was supported in part by funding from the Digital Research Alliance of Canada as part of the DRI EDIA Champions project. The workshop sessions were co-organized and promoted with Let's Talk Science Canada. J.G. acknowledges the NSERC CREATE in Quantum Computing Program (Grant Number  543245), the Wendy Diane Esdale Graduate Scholarship, the Department of Chemistry Summer Graduate Award-Research Assistantship, and the David McGillivray Scholarship in Science from the University of Victoria.

This research was undertaken also in part to funding from the Canada Research Chairs Program (CRC-2021-00257), and supported in part by the Natural Sciences and Engineering Research Council of Canada (NSERC) under grants RGPIN-2023-05510 and DGECR-2023-00026. 

\bibliography{sn-bibliography}

\end{document}